\documentclass[11pt]{article}
\usepackage{amsfonts}
\usepackage{amssymb}
\usepackage{amsmath}
\def\@begintheorem#1#2{\par\bgroup{\sc #1\ #2. }\it\ignorespaces}
\def\@opargbegintheorem#1#2#3{\par\bgroup{\sc #1\ #2\ (#3). } \it\ignorespaces}
\def\@endtheorem{\egroup}
\newtheorem{theorem}{Theorem}[section]
\newtheorem{corollary}[theorem]{Corollary}
\newtheorem{proposition}[theorem]{Proposition}
\newcommand{\bt}[1]{\begin{theorem}\label{#1}}
\newcommand{\bc}[1]{\begin{corollary}\label{#1}}
\newcommand{\bp}[1]{\begin{proposition}\label{#1}}
\newcommand{\et}{\end{theorem}}
\newcommand{\ec}{\end{corollary}}
\newcommand{\ep}{\end{proposition}}

\def\R{\mathbb{R}}
\def \P {{{\cal P}}}
\def \conv {{\rm conv}}
\def \p {\psi_{n,n}}
\def \f {\psi_{n}}
\def \l {\langle}
\def \r {\rangle}
\def \o {\otimes}
\def \t {{^\intercal}}

\begin{document}

\title{\bf Two graph isomorphism polytopes}
\author{Shmuel Onn}

\date{January 11, 2008}
\maketitle

\begin{abstract}
The convex hull $\p$ of certain $(n!)^2$ tensors was considered recently in
connection with graph isomorphism. We consider the convex hull $\f$
of the $n!$ diagonals among these tensors. We show: 1. The polytope $\f$ is a face of $\p$.
2. Deciding if a graph $G$ has a subgraph isomorphic to $H$ reduces to
optimization over $\f$. 3. Optimization over $\f$ reduces to optimization over $\p$.
In particular, this implies that the subgraph isomorphism problem reduces
to optimization over $\p$.
\vskip.2cm
\noindent
{\em AMS Subject Classification:}
05A, 15A, 51M, 52A, 52B, 52C, 62H, 68Q, 68R, 68U, 68W, 90B, 90C
\end{abstract}

\section{Introduction}

Let $\P_n$ be the set of $n\times n$ permutation matrices
and consider the following two polytopes,
$$\f\ :=\ \conv\{P\o P\ :\ P\in \P_n\}\,,\quad\quad
\p\ :=\ \conv\{P\o Q\ :\ P,Q\in \P_n\}\ .$$
The polytope $\p$ was considered recently in \cite{Fri}
in connection with the graph isomorphism problem.
Note that $\f$ and $\p$ have $n!$ and $(n!)^2$ vertices respectively.
In this short note we show:
\begin{enumerate}
\item The polytope $\f$ is a face of the polytope $\p$.
\item Deciding if a graph $G$ has a subgraph isomorphic to
a graph $H$ reduces to optimization over $\f$.
\item Optimization over $\f$ reduces to optimization over $\p$.
\end{enumerate}
In particular, this implies a result of \cite{Fri} that subgraph
isomorphism reduces to optimization over $\p$.

So if $P\neq NP$ then optimization and separation over $\f$ and hence over
$\p$ cannot be done in polynomial time and a compact inequality description
of $\f$ and hence of $\p$ cannot be determined.

Deciding if $G$ has a subgraph that is isomorphic to $H$ can also be reduced to
optimization over a related polytope $\phi_n$ defined as follows.
Each permutation $\sigma$ of the vertices of the complete graph $K_n$
naturally induces a permutation $\Sigma$ of its edges by $\Sigma(\{i,j\}):=\{\sigma(i),\sigma(j)\}$.
Then $\phi_n$ is defined as the convex hull of all ${n\choose 2}\times {n\choose 2}$
permutation matrices of induced permutations $\Sigma$. This polytope and a broader
class of so-called {\em Young polytopes} have been studied in \cite{Onn}.
In particular, therein it was shown that the graph of $\phi_n$ is complete,
so pivoting algorithms cannot be exploited for optimization over this polytope.
It is an interesting question whether $\f$ and $\phi_n$,
having $n!$ vertices each, are isomorphic.

\section{Statements}

Define bilinear forms on $\R^{n\times n}$ and on $\R^{n\times n}\o\R^{n\times n}$
(note the shuffled indexation on the right) by
$$\l A,B\r:=\sum_{i,j}A_{i,j}B_{i,j}\,,\quad\quad
\l X , Y\r\ :=\ \sum_{i,j,s,t}X_{i,s,j,t}Y_{i,j,s,t}\ .$$
Let $I$ be the $n\times n$ identity matrix
and for a graph $G$ let $A_G$ be its adjacency matrix. We show:
\bt{T1}
The polytope $\f$ is a face of $\p$ given by
$$\f\quad =\quad \p\ \cap\ \{X\ :\ \l I\o I,X\r\ =\ n\}\ .$$
\et
\bt{T2}
Let $G$ and $H$ be two graphs on $n$ vertices with $m$ the number of edges of $H$. Then
$$\max\{\l A_G\o A_H,X\r\ :\ X\in\f\}\quad \leq \quad 2m$$
with equality if and only if $G$ has a subgraph that is isomorphic to $H$.
\et
\bt{T3}
Let $W=(W_{i,s,j,t})$ be any tensor and let $w:=2n^2\max|W_{i,s,j,t}|$. Then
$$\max\{\l W,X\r\ :\ X\in\f\}\quad =\quad
\max\{\l W+w I\o I,X\r\ :\ X\in\p\}\ -\ nw\ .$$
\et
Combining Theorems \ref{T2} and \ref{T3} with $W=A_G\o A_H$ and $w=n^2$
(sufficing since $W\geq 0$, as is clear from the proof of Theorem \ref{T3} below),
we get the following somewhat tighter form of a result of \cite{Fri}.
\bc{C}
Let $G$ and $H$ be two graphs on $n$ vertices with $m$ the number of edges of $H$. Then
$$\max\{\l A_G\o A_H + nI\o nI,X\r\ :\ X\in\p\}\quad \leq \quad 2m\ +\ n^3$$
with equality if and only if $G$ has a subgraph that is isomorphic to $H$.
\ec

\section{Proofs}

We record the following statement that follows directly
from the definitions of the bilinear forms above.
\bp{simple}
For any two simple tensors $X=A\o B$ and $Y=P\o Q$ we have
$$\l X, Y\r\ =\ \l A\o B, P\o Q\r\ =\
\sum_{i,j,s,t}A_{i,s}B_{j,t}P_{i,j}Q_{s,t}\ =\ \l PBQ^\t, A\r\ .$$
\ep
{\bf \em Proof of Theorem \ref{T1}.}$\ $
For every $P,Q\in\P_n$, the matrix $PIQ^\t$ is a permutation matrix,
with $PIQ^\t=I$ if and only if $P=Q$. It follows that for every two
distinct $P,Q\in\P_n$ we have
\begin{equation}\label{e1}
\l I\o I,P\o Q \r\ =\ \l PIQ^\t,I \r\ \leq \ n-1\ <\
n\ =\ \l PIP^\t,I \r\ =\ \l I\o I,P\o P \r\ .\quad\ \ \blacksquare
\end{equation}

\vskip.2cm\noindent
{\bf \em Proof of Theorem \ref{T2}.}$\ $
For any $P\in\P_n$, the matrix  $PA_HP^\t$ is the adjacency matrix of
the permutation of $H$ by $P$. So $\l PA_HP^\t, A_G\r\leq 2m$ with equality
if and only if $H$ is isomorphic via $P$ to a subgraph of $G$.
Since the maximum of a linear form over a polytope
is attained at a vertex we get
\begin{eqnarray*}
\max\{\l A_G\o A_H,X\r\ :\ X\in\f\}
& = & \max\{\l A_G\o A_H, P\o P\r\ :\ P\in\P_n\} \\
& = & \max\{ \l PA_HP^\t, A_G\r\ :\ P\in\P_n\}\ \ \leq\ \ 2m
\end{eqnarray*}
with the last inequality holding with equality
if and only if $G$ has a subgraph isomorphic to $H$.
$\quad\ \ \blacksquare$

\vskip.2cm\noindent
{\bf \em Proof of Theorem \ref{T3}.}$\ \ $
For every $P,Q\in\P_n$, the tensor $P\o Q=(P_{i,j}Q_{s,t})$ has $n^2$
entries that are equal to $1$ and all other entries equal to $0$,
and therefore $-{1\over2}w\leq\l W,P\o Q\r\leq{1\over2} w$.
Combining this with inequality (\ref{e1}) we see that for every
two distinct $P,Q\in\P_n$ we have
\begin{eqnarray*}
\l W+ wI\o I,P\o Q \r
& = & \l W,P\o Q\r \ +\ w \l I\o I,P\o Q\r \\
& \leq & {1\over2}w\ +\ (n-1)w\ \ =\ \ -{1\over2}w\ +\ nw  \\
& \leq & \l W,P\o P\r \ +\ w \l I\o I,P\o P\r \ \ =\ \ \l W+ wI\o I,P\o P \r \ .
\end{eqnarray*}
Since the maximum of a linear form over a polytope
is attained at a vertex we obtain the inequality
\begin{eqnarray*}
\max\{\l W+wI\o I,X\r\ :\ X\in\p\}
& = & \max\{\l W+wI\o I,P\o Q\r\ :\ P,Q\in\P_n\} \\
& = & \max\{\l W+wI\o I, P\o P\r\ :\ P\in\P_n\} \\
& = & \max\{\l W, P\o P\r\ :\ P\in\P_n\}\ +\ nw \\
& = & \max\{\l W, X\r\ :\ X\in\f\}\ +\ nw
\ .\quad\ \ \blacksquare
\end{eqnarray*}

\vskip.3cm\noindent {\small Shmuel Onn}\newline
\emph{Technion - Israel Institute of Technology, 32000 Haifa, Israel}\newline
\emph{email: onn{\small @}ie.technion.ac.il},
\ \ \emph{http://ie.technion.ac.il/{\small $\sim$onn}}

\end{document}